\documentclass[twocolumn,prl,aps,showpacs]{revtex4}
\usepackage{amssymb}
\usepackage{epsfig,amsmath}
\usepackage{graphicx}
\usepackage{dcolumn}
\usepackage{bm}

\newcommand{\beq}{\begin{equation}}
\newcommand{\eeq}{\end{equation}}
\newcommand{\beqa}{\begin{eqnarray}}
\newcommand{\eeqa}{\end{eqnarray}}

\newcommand{\la}{\langle}
\newcommand{\ra}{\rangle}

\def\nat#1{{ Nature} {\bf#1}}

\def\npho#1{{Nature\ Phot.} {\bf#1}}

\def\ol#1{{ Opt.\ Lett.} {\bf#1}}
\def\pla#1{{ Phys.\ Lett. A\/} {\bf#1}}

\def\pra#1{{ Phys.\ Rev. A\/} {\bf#1}}

\def\prl#1{{ Phys.\ Rev.\ Lett.} {\bf#1}}

\def\sci#1{{ Science} {\bf#1}}
\def\rmp#1{{ Rev. \ Mod. \ Phys.} {\bf#1}}

\begin{document}

\title{Entanglement is Sometimes Enough}
\author{Xiao-Feng Qian}
\email{xfqian@pas.rochester.edu}
\author{J.H. Eberly}
\affiliation{Rochester Theory Center and the Department of Physics
\&
Astronomy\\
University of Rochester, Rochester, New York 14627}
\date{\today }

\begin{abstract}
For many decades the word ``entanglement" has been firmly attached
to the world of quantum mechanics. So is the phrase ``Bell
violation". Here we show, without contradicting quantum mechanics,
that classical non-deterministic fields also provide a natural basis
for entanglement and Bell analyses. Surprisingly, such fields are
not eliminated by the Clauser-Horne-Shimony-Holt Bell violation test
as viable alternatives to quantum theory. An experimental setup for
verification is proposed.
\end{abstract}


\maketitle



Without contradicting quantum mechanics, we show that some classical
field theories can provide entanglement and predict Bell violation
while also complying with all three of Shimony's criteria
\cite{Shimony} for a viable theory of Nature. No viable field theory
candidates other than quantum mechanics have previously survived the
Bell violation test. The incentive to demonstrate that classical
fields can have all the desired properties is provided by a logical
gap in the following remark made by John Bell \cite{BellScience} in
1972:  ``It can indeed be shown that the quantum mechanical
correlations cannot be reproduced by a hidden variables theory even
if one allows a `local' sort of indeterminism. .... This would not
work."

But it does work and what Bell did not anticipate, now well
accepted, was that classical fields allow intrinsic entanglement.
The remaining gap has been filled by the recognition
\cite{Simon-etal, Qian-EberlyOL} that examples exist among fields
that are intrinsically entangled indeterministically.  These fields
allow satisfaction of all three criteria of Shimony for a viable
theory. These criteria accept the randomness of the micro-world by
emphasizing the need for indeterminism: ``(I) In any state of a
physical system S there are some eventualities which have indefinite
truth values. (II) If an operation is performed which forces an
eventuality with indefinite truth value to achieve definiteness ...
the outcome is a matter of chance. (III) There are `entangled
systems' (in Schr\"odinger's phrase) which have the property that
they constitute a composite system in a pure state, while neither of
them separately is in a pure state." By eventualities Shimony just
means measurement outcomes.


Bell's motivations were clear and are an important guide. We will
exhibit a classical physical theory that is intrinsically
non-deterministic and allows, contrary to Bell's assertion, both
derivations and violations of the same Bell Inequalities. This
apparently self-contradictory possibility, which arises when
addressing field (wave) rather than particle aspects of natural
phenomena, will be resolved below. The category of theories open for
study has a potentially large number of members, and at least one
very well known member, namely optical partial coherence theory
\cite{Brosseau, Wolf}. We use this as our example. It is compatible
with the conclusions of the Bell violation experiments because Bell
violation occurs within it in exactly the same way and to exactly
the same degree as in quantum theory. We demonstrate this in the
context of a two-party CHSH Bell Inequality \cite{CHSH} that
embraces entanglement (without quantization), and we show that it is
subject to experimental verification. We are apparently dealing with
a domain where characteristics labeled classical and quantum have
not yet been definitively separated.


Attending to Shimony's validity conditions (I)-(III), the need for
indefinite truth values is automatically met in the indeterministic
theories that concerned Bell. In the ordinary classical optical
coherence theory of partial polarization  \cite{Brosseau, Wolf} one
interprets a light field's physical variables indeterministically.
As a concrete system one can think of a beam of thermal light, in
which the value of the ``optical field" is unpredictable. The real
and in principle observable optical field of such a light beam is
simply the space-time dependent electric field vector \beq
\label{Efield} \vec E(r,t) = \hat h E_h(r,t) + \hat v E_v(r,t), \eeq
where $\hat h$, $\hat v$ are orthogonal and arbitrarily oriented
``horizontal" and ``vertical" polarization directions.  The
corresponding components $E_h(r,t) = \sum_nh_n(t)\phi_n(r)$ and
$E_v(r,t) = \sum_lv_l(t)\phi_l(r)$ of the optical field are elements
in an abstract stochastic function space \cite{circpolzn}. Here the
$\phi_n(r)$ are orthonormal spatial mode functions with $\int
dr\phi^{*}_n(r)\phi_l(r) = \delta_{n,l}$, and $h_n(t)$ and $v_l(t)$
represent stochastic random coefficients whose origin we can assign
to distant and unknowable dipole radiators.

In thermal the value of $\vec E$ is unpredictable at any time, so
criterion (I) about indefinite truth values is satisfied. When the
field is appropriately detected, only one value is recorded, and so
a definite truth value is then obtained, satisfying criterion (II).
No one would think that this picture of the optical field
(\ref{Efield}) implies that it is quantized, and it is not, but a
non-deterministic viewpoint is employed to extract predictions from
the randomly unknown ensemble of field potentialities. This is
conventionally done via observable field correlation functions.

Criterion III, the need for entanglement, can be thought difficult.
When he introduced the word ``entanglement" in 1935, Schr\"odinger
unfortunately created the nearly indelible impression that
entanglement is identified  exclusively with quantum mechanics.
However, the truth is, as Schr\"odinger clearly knew \cite{Schr},
that entanglement had already been part of an exhaustive study of
integral equations in Hilbert space in 1907 by Schmidt
\cite{Schmidt} when Schr\"odinger was just a teenager, two decades
before quantum mechanics even existed.


As it turns out, condition III, the existence of entanglement, is
readily associated with partial polarization in optics. This has
been recently emphasized by Simon, et al. ~\cite{Simon-etal} and
Qian and Eberly \cite{Qian-EberlyOL} for non-deterministic optical
fields \cite{alternatives}. Entanglement requires sums of tensor
products of vectors from at least two vector spaces, and optical
field (\ref{Efield}) is clearly such a sum, where the vectors are
associated with the ``lab space" of $\hat h$ and $\hat v$ on the one
hand and on the other hand with the statistical infinite-dimensional
continuous Hilbert space of the components $E_h$ and $E_v$.

In very special cases there is a direction $\hat u$ (a particular
linear superposition of $\hat h$ and $\hat v$) that turns
(\ref{Efield}) into $\vec E = \hat u F(r,t)$. This allows the two
spaces to be separated (factored), and such a separable field is
obviously completely polarized (in the direction $\hat u$). Any
non-factorable form for $\vec E$ represents partial polarization,
and a finite degree of (fully classical) entanglement
\cite{Qian-EberlyOL}. We will identify $I = \la E_hE_h \ra + \la
E_vE_v \ra$ as the light intensity. Here the angle bracket $\la
...\ra $ denotes a combination of an ensemble average for the random
coefficients, and a spatial average for the mode functions because
we are considering the character of an entire light beam instead of
specific modes or photons.

Thermal light is understood completely both classically and quantum
mechanically and in both domains is probabilistically characterized
as completely uncorrelated, Gaussian, and statistically stationary,
with the consequence that $E_h$ and $E_v$ have zero statistical
overlap: $\la E_hE_v \ra = 0$ with equal magnitudes. The classical
(and quantum) degree of polarization of such a field is zero and the
degree of entanglement is maximal. In fact a purely thermal field is
a classical Bell state of the two ``parties" in lab space and
function space.

To encourage the picture of the optical field as a state in a
product tensor space, which it clearly is, we will adopt Dirac-type
notation for the vectors: $\vec E \to |{\bf E}\ra$, $\hat h \to
|h\ra$, etc., where we use boldface to emphasize the ``bi-vector"
two-space character of the field: \beq |{\bf E}\ra = |h\ra \otimes
|E_h\ra + |v\ra \otimes |E_v\ra. \eeq The theory is thus built on
superposable states (wave fields), but without any implication that
quantization has been imposed.  For the general partially coherent
case, then, we can write the normalized field $|{\bf e}\ra \equiv
|{\bf E}\ra/\sqrt{I}$ as: \beq \label{e-def} |{\bf e}\ra =
\kappa_{1}|u_1\ra \otimes |f_1\ra + \kappa_{2}|u_2\ra \otimes
|f_2\ra, \eeq where $\la u_j |u_k\ra = \la f_j |f_k \ra =
\delta_{jk}$, and $\kappa_{1}$ and $\kappa_{2}$ are normalization
coefficients, both equal to $1/\sqrt{2}$ in the exactly thermal
case. For an arbitrary field state such a decomposition is
guaranteed by the Schmidt theorem (see \cite{Qian-EberlyOL}).


Bell's agenda \cite{BellSpeakable} was to ascertain measurement
probabilities and correlations between vectors in separate vector
spaces, and polarization vectors of different photons have been used
most frequently for the observations. Under obvious conditions on
observability, and independent of the possible existence of hidden
control parameters, various well-known ``Bell Inequalities" serve to
constrain the range of these correlations. The CHSH
(Clauser-Horne-Shimony-Holt) Inequality \cite{CHSH} is the most
useful of these inequalities and has been employed repeatedly
\cite{tests}. As Gisin \cite{Gisin} has observed, it can be
straightforwardly violated in correlation measurements made on any
quantum mechanical pair of vectors in a state of the same form as
(\ref{e-def}).

The new result presented here is that the same violation should also
be expected classically. Any classical violation contradicts Bell's
conclusion that quantum theory must be the explanation for any
violation. The fact is that in all cases, quantum and classical, it
is entanglement that provides the violation.

\begin{figure}[t!]
\includegraphics[width= 6 cm]{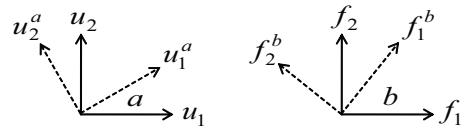}
\caption{\footnotesize{Rotations of lab and function space
coordinates through angles $a$ and $b$, respectively.}}
\label{angles}
\end{figure}

We replay a derivation of the CHSH inequality in the supplemental
material, and only sketch the main features here. A standard
Bell-CHSH approach is based on correlations observed in various
rotated basis frames. For the case of light beam (\ref{e-def}),
rotations such as shown in Fig. \ref{angles} for lab and function
spaces can be denoted as \beqa &&|{u}_{1}^{a}\rangle =\cos
a|{u}_{1}\rangle -\sin a|{u}_{2}\ra \quad {\rm
and} \notag \\
&& |{u}_{2}^{a}\ra =\sin a|{u}_{1}\ra +\cos a|{u}_{2}\ra. \notag
\eeqa To characterize the beam (\ref{e-def}) in lab space, one can
make a projection in any rotated basis $|{u}_{k}^{a}\ra$, $k=1,2$,
so that one has $|{\bf e}_{k}^{a}\ra \equiv |{u}_{k}^{a}\ra
\la{u}_{k}^{a}| {\bf e}\ra$ which leaves the field only in the
$|{u}_{k}^{a}\ra$ lab direction. Then one can always obtain the
fraction of intensity in this component as
$$P_{k}(a) = |\la{u}_{k}^{a}| {\bf e}\ra|^2 = \la {\bf e}|u_{k}^{a}\ra \la
u_{k}^{a}|{\bf e}\ra \equiv \la {\bf e}|
\overleftrightarrow{p}_{k}^{a}|{\bf e}\ra.$$ The dimensionless
fractions $P_{k}(a)$ can obviously be interpreted as probabilities
and they satisfy the natural relation $P_{1}(a) + P_{2}(a) = 1$. To
capture the individual projections $P_{1}(a)$ and $P_{2}(a)$ in lab
space we define a new outer product
\begin{equation} \label{measurementA}
\overleftrightarrow{A}_a = \overleftrightarrow{p}_{1}^{a} -
\overleftrightarrow{ p}_{2}^{a},
\end{equation}
which fully characterizes the beam (\ref{e-def}) in lab space. Its
average $A(a)=\la {\bf e}| \overleftrightarrow{A}_a|{\bf e}\ra$ is a
real number between $-1$ and $1$, and it fully determines both
$P_{1}(a)$ and $P_{2}(a)$ at the same time, through use of $P_{1}(a)
+ P_{2}(a) = 1$. Similarly one can also characterize the beam in the
function space by defining the outer product
$\overleftrightarrow{B}_b = \overleftrightarrow{p}_{1}^{b} -
\overleftrightarrow{p}_{2}^{b}$, where
$\overleftrightarrow{p}_{l}^{b} = |f_{l}^{b}\rangle \langle
f_{l}^{b}|$ with $l=1,2$. Then its average value $B(b) = \la {\bf
e}| \overleftrightarrow{B}_b|{\bf e}\ra$ is also bounded by $-1$ and
$1$. Finally, the measurement outcome correlation between the lab
and function spaces can be written
\begin{equation}
C(a,b)=\la {\bf e}|\overleftrightarrow{A}_a\otimes
\overleftrightarrow{B}_b|{\bf e}\ra, \label{correlation1}
\end{equation}
which is a combination of 4 joint probabilities \beq \label{Pklab}
P_{kl}(a,b) = \la {\bf e}|\Big(|{u}_{k}^{a}\ra |f_{l}^{b}\ra \la
f_{l}^{b}|\la {u}_{k}^{a}|\Big)|{\bf e}\ra, \eeq with $k,l = 1,2 $.
If one defines  $S \equiv C(a,b) - C(a,b') + C(a',b) + C(a',b')$,
where $a,\ a',\ b,\ b'$ are arbitrary rotation angles, then one can
obtain the CHSH result \beq \label{S ineq} -2 \leq S \leq 2. \eeq We
remark that the lab and function spaces, based on which all the
correlations are analyzed, belong to the same entity (i.e., the
thermal light beam), so the usual locality assumption is replaced
here by the statistical independence assumption in the derivation of
the above inequality (see supplemental material). This replacement
is also made in quantum analyses of Bell inequalities for hybrid
entanglement (see for example \cite{Karimi-etal, Leuchs-etal} and
references therein).

Note that, for an ideal thermal beam, classical optics gives $C(a,b)
= \cos 2(a-b)$. This special case paradoxically tells us that there
must be a violation of the CHSH Inequality for classically
non-deterministic beams, even though only classical results are used
to obtain the inequality. Obviously, this requires comment below.

Going further, for the general field (\ref{e-def}), the joint
probabilities in terms of rotated angles $a$ and $b$ can all be
calculated and have familiar values in classical statistical optics.
The result is an explicit formula for any classical indeterministic
$S$: \beqa \label{CHSH2}
S &=& 2\kappa_{1}\kappa_{2}[\sin 2a(\sin 2b-\sin 2b') \notag \\
&+& \sin2a'(\sin 2b+\sin 2b')] \notag \\
&+& \cos 2a(\cos 2b-\cos 2b') \notag \\
&+& \cos 2a'(\cos 2b+\cos 2b'). \eeqa One can choose the angles
freely, and a useful choice is $a=0,\ a'=\pi/4,\ b =\pi/8,
b'=3\pi/8$. Then one quickly finds that $S$ takes its maximum value
for $\kappa_{1} = \kappa_{2} = 1/\sqrt{2}$, and then $S = \sqrt{2}(1
+ 2\kappa_{1}\kappa_{2}) \to 2\sqrt 2$, which obviously violates the
inequality $|S| \leq 2$.   The only difference to a familiar quantum
derivation is that $A(a)$ and $B(b)$ both lie anywhere in the
continuum between $-1$ and $+1$, rather than taking discrete values
such as $\pm 1$. Here we have no quantum particles to be detected or
counted, but a statistical light beam with variable intensity
between its components.

\begin{figure}[t!]
\includegraphics[width=8 cm]{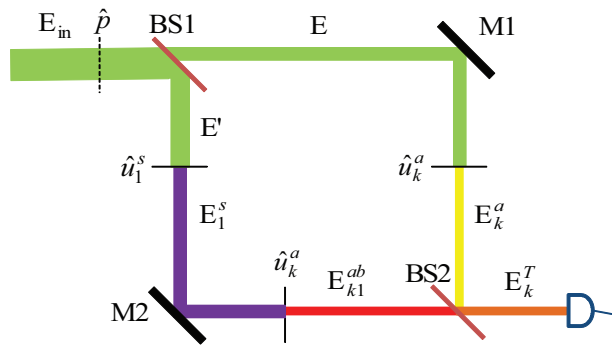}
\caption{Schematic experimental setup. The polarizer $\hat{p}$ is
inserted temporarily to measure the values of $\kappa_1$, $\kappa_2$
and the directions of $\hat{u}_1$, $\hat{u}_2$ in equation
(\ref{e-def}), and is then removed. The incoming partially polarized
master beam $|{\bf E}_{in}\ra$ (in green) is directed to an ordinary
beam splitter (BS1). The transmitted beam $|{\bf E}\ra$, via mirror
(M1), passes through a polarizer $\hat{u}_{k}^{a}$. The reflected
beam $|{\bf E}'\ra$ passes through a polarizer $\hat{u}_{1}^{s}$ and
via mirror (M2) passes through polarizer $\hat{u}_{k}^{a}$. The two
beams are combined by a 50:50 beam splitter (BS2) and detected. The
width and color of a beam schematically indicate intensity and
statistical characteristic respectively.} \label{setup}
\end{figure}

We now describe the experiment sketched in Fig. \ref{setup} to test
these theoretical predictions with a non-deterministic classical
light beam. The experiment is designed to determine the correlation
function $C(a,b)$ through the joint probabilities (\ref{Pklab}) by
measuring various intensities. Measurements and projections in lab
space can be achieved by passing the light beam through polarizers
placed at desired angles. Photon detection is clearly not called for
and even a calorimeter could be used to measure the intensities of
the input and output beams, and thus determine the outcome
probabilities.

It is obvious that the two sub-beams $|{\bf E}\rangle$ and $|{\bf
E}'\rangle$ inherit the statistical properties of the master beam
and thus both can be expressed as Eq. (\ref{e-def}), with
corresponding intensities $I$ and $I'$. An unimportant $i$ phase
comes from the beam splitter. To determine the joint probabilities
$P_{kl}(a,b)$ of the test beam $|{\bf E}\rangle$, the first natural
step is to project the field in the lab space onto the basis
$|u_{k}^{a}\rangle $. This can be realized by a polarizer
$\hat{u}_{k}^{a}$, as shown in the figure, allowing only the
$|u_{k}^{a}\ra$ component to pass. Then the corresponding
transmitted beam becomes
\begin{equation}
|{\bf E}_{k}^{a}\ra =\sqrt{I_{k}^{a}}|u _{k}^{a}\ra
(A_{k1}|f_{1}^{b}\ra + A_{k2}|f_{2}^{b}\ra),
\end{equation}
where $I_{k}^{a}$ is the intensity, and $A_{kl}$ with $k,l=1,2$, are
normalized amplitude coefficients which relate to joint
probabilities in an obvious way:
$P_{kl}(a,b)=I_{k}^{a}|A_{kl}|^{2}/I$. One sees that the intensities
$I$ and $I_{k}^{a}$ can be measured directly but not the
coefficients $A_{kl}$.

The statistically identical auxiliary beam $|{\bf E}'\ra$ is used to
help determine these coefficients. One needs to produce a beam that
carries only one of the two components $|f_{1}^{b}\rangle$,
$|f_{2}^{b}\rangle$. However, there is no technology for rotation in
function space, i.e., for projecting a non-deterministic field onto
an arbitrary direction in the continuously infinite-dimensional
function space. We have solved this problem with an indirect
measurement setup by passing the beam through a polarizer
$\hat{u}_{1}^{s}$ that passes only the special lab space vector
$|u_{1}^{s}\ra$, rotated by angle $s$ from the initial basis
$|u_{1}\ra$, where the angle $s$ is chosen so that the other
statistical component $|f_{2}^{b}\ra$ is stripped off. The
transmitted beam $|{\bf E}_{1}^{s}\ra$ has only the $|f_{1}^{b}\ra$
component in the function space, i.e., $|{\bf E}_{1}^{s}\ra =
i\sqrt{I_{1}^{s}}|{u}_{1}^{s}\ra |f_{1}^{b}\ra$. Here $I_{1}^{s}$ is
the corresponding intensity and the special stripping angle $s$ can
be determined by the relation
\begin{equation}
\tan s = (\kappa_{1}/\kappa_{2})\tan b.
\end{equation}

The function-space-oriented beam $|{\bf E}_{1}^{s}\rangle $ is then
sent through polarizer $\hat{u}_{k}^{a}$ that transmits only the
$|{u}_{k}^{a}\ra $ component in the lab space and becomes $|{\bf
E}_{k1}^{ab}\ra =|{u}_{k}^{a}\ra \la {u}_{k}^{a}|{\bf E}_{1}^{s}\ra
=i\sqrt{I_{k1}^{ab}}|{u} _{k}^{a}\ra |f_{1}^{b}\ra$, where
$I_{k1}^{ab}$ is the corresponding intensity. Finally, as shown in
Fig. \ref{setup}, the beams $|{\bf E}_{k}^{a}\ra$ and $|{\bf
E}_{k1}^{ab}\ra$ are combined by a 50:50 beam splitter which yields
the outcome beam as $|{\bf E}_{k}^{T}\ra = (|{\bf E}_{k1}^{ab}\ra +
i|{\bf E}_{k}^{a}\ra)/\sqrt{2}$. Then the intensity $I_{k}^{T}$ of
this outcome beam can be easily expressed in terms of the
coefficients $A_{kl}$. Some simple arithmetic will immediately
provide the joint probabilities $P_{kl}(a,b)$ in terms of various
intensities
\begin{eqnarray}
P_{k1}(a,b) &=&(2I_{k}^{T}-I_{k1}^{ab}-I_{k}^{a})^{2}/4II_{k1}^{ab},
\notag \\
P_{k2}(a,b)
&=&I_{k}^{a}/I-(2I_{k}^{T}-I_{k1}^{ab}-I_{k}^{a})^{2}/4II_{k1}^{ab},
\end{eqnarray}
where $k=1,2$. Therefore the joint probabilities can be obtained by
measuring four different intensities, $I$, $I_{k}^{a}$,
$I_{k1}^{ab}$ and $I_{k}^{T}$. One can immediately achieve the
correlation function $C(a,b)$ as defined in (\ref{correlation1}),
and consequently the value of $S$ by carrying out three more sets of
experiments with different combinations of polarizer direction
setups to achieve the remaining three correlations $C(a,b')$,
$C(a',b)$, and $C(a',b')$.


In deriving an inequality and then showing that it can be violated
within the same framework as its derivation, one is certain to be
making a mis-step. The mis-step is of course the main point. What we
have done that leads to the apparent contradiction is to make the
statistical independence assumption contained in all CHSH Inequality
derivations. Because of the independence of methods used to register
the vectors in the two distinct vector spaces, it is always assumed
that all possible connections, beyond those originating with the
$\{\lambda\}$ distribution, have been excluded. But the presence of
entanglement supplies another connection, one that by its nature
bridges the two vector spaces. This exposes the fact that what is
normally regarded as quantum behavior (indeterminism plus Bell
violation) is not restricted to quantum contexts. Our results allow
one to say that Bell violation can be explained without quantum
mechanics, but not without entanglement.

In conclusion we have provided a test of the uniqueness of quantum
theory. We have shown that the statistical theory of optical
coherence embodies the three ``viable theory" criteria of Shimony,
and it allows both the standard derivation of the CHSH Bell
Inequality and the violation of it. This contradiction is naturally
explained, as in the preceding sentences, and is interesting because
it refutes the understanding by Bell \cite{BellScience}, accepted
generally since, that correlations violating a Bell Inequality
cannot originate in other than quantum states. Our results thus
clarify two long-standing and widely held mis-impressions: that
indeterministic entanglement is unique to quantum mechanics, and
that quantum mechanics is unique to violate Bell inequalities.
Although our discussion was based on the example of
non-deterministic light fields, it seems clear that the analysis can
be extended to general stochastic wave theories.

The authors acknowledge helpful discussions over several years with
colleagues including A. Aspect, A.F. Abouraddy, J.A. Bergou, C.
Broadbent, L. Davidovich, J. Dressel, E. Giacobino, J.C. Howell,
D.F.V. James, A.N. Jordan, H.J. Kimble, G. Leuchs, P.W. Milonni, M.
Segev, R.J.C. Spreeuw, J.P. Woerdman and T. Yu, as well as financial
support from NSF PHY-0855701 and NSF PHY-1203931.


\section*{Supplemental Material}

Review of CHSH derivation. Due to the non-deterministic
characteristics (of the thermal light field), Bell analysis allows
that the average measurement outcomes of one space could be
influenced by some unknown parameters as well as hidden control
variables. We follow tradition and label these contextual unknowns
collectively by a single multi-dimensional parameter $\{\lambda\}$
with an overall unspecified distribution $\rho(\{\lambda\})$.
Therefore the measurement results can be rewritten in terms of
probabilities admitting such cross-dependent conditional
possibilities:
\begin{equation*}
A(a) \to A(a,\{\lambda\}|B) = P_{1}(a,\{\lambda\} |B) -
P_{2}(a,\{\lambda\}|B),
\end{equation*}
and similarly
\begin{equation*}
B(b) \to B(b,\{\lambda\}|A) = P_{1}(b,\{\lambda\}|A) -
P_{2}(b,\{\lambda\} |A).
\end{equation*}

Probability measurement setups in one space are independent of
measurement setups in the other space. This allows a measurement
outcome statistical independence assumption, so the probabilities
reduce to the simpler forms
\begin{equation}  \label{assumption}
P_{k}(a,\{\lambda\} |B) = P_{k}(a,\{\lambda\} ),\
P_{l}(b,\{\lambda\} |A) = P_{l}(b,\{\lambda\} ).  \notag
\end{equation}
This does not exclude correlations between the measurements in the
two spaces, i.e., the outcomes in both spaces may still be related
because of $\{\lambda\}$. However, as usual, it says that joint
$a-b$ probabilities become products of individual probabilities:
$P_{kl}(a,b,\{\lambda\}) = P_{k}(a,\{\lambda\})
P_{l}(b,\{\lambda\})$. Consequently the correlation function (5) in
the Letter, for arbitrary angles $a$ and $b$, is equivalent in the
usual way to
\begin{equation}
C(a,b) = \int A(a,\{\lambda\} )B(b,\{\lambda\} )\rho
(\{\lambda\})d\{\lambda\}.  \notag  \label{correlation2}
\end{equation}

Then one can follow the standard CHSH procedure and find
\begin{eqnarray*}
S &=&C(a,b)-C(a,b')+C(a',b)+C(a',b') \\
&=&\int A(a,\{\lambda \})[B(b,\{\lambda \})-B(b',\{\lambda \})] \\
&&+A(a',\{\lambda \})[B(b,\{\lambda \})+B(b',\{\lambda \})]\rho
(\{\lambda \})d\{\lambda \}.
\end{eqnarray*}
From the fact that any measurement results $A(a,\{\lambda \})$ and
$B(a,\{\lambda \})$ lie between the values $-1$ and $1$, the above
expression of $S$ straightforwardly obeys the inequality $|S|\leq 2$
for any $a$, $a'$, $b$, $ b'$. This concludes the derivation of CHSH
inequality.


\begin{thebibliography}{99}
\bibitem{circpolzn} We will continue to pretend that all the functions and
vectors are real, but allowing them to be complex or allowing the
unit vectors to be time-dependent doesn't change anything important
(see, e.g., \cite{Brosseau, Wolf}).
\end{thebibliography}
\end{document}